\documentclass{ws-ijmpa}
\usepackage[super,compress]{cite}
\usepackage{graphicx}

\usepackage{amsmath, amssymb, amsbsy, amsfonts, dsfont}
\usepackage{srcltx, color}

\usepackage{makecell}

\usepackage[table]{xcolor}
\definecolor{lightgray}{gray}{0.9}

\begin{document}
\markboth{E.T. Kipreos}{Assessing experimental support for differential simultaneity}

\catchline{}{}{}{}{}
%
%%%%%%%%%%%%%%%%%%%%%%%%%%%%%%%%%%%%%%%%%%%%%%%%%%%%%%%%%%%%%%%%%%%%

\title{Assessment of the experimental support for differential simultaneity}

\author{Edward T. Kipreos}

\address{University of Georgia, 136 Cedar Street,\\
Athens, Georgia 30602, USA\\
ekipreos@uga.edu}

%\author{Second Author}

%\address{Group, Laboratory, Address\\
%City, State ZIP/Zone, Country\\
%second\_author@domain\_name}

\maketitle

\begin{history}
%\received{Day Month Year}
%\revised{Day Month Year}
\end{history}

\begin{abstract}
The Lorentz transformation describes differential simultaneity, which reflects the offsetting of time with distance between reference frames.  Differential simultaneity is essential for Lorentz invariance.  Here, the current experimental support for differential simultaneity is assessed.  Differential simultaneity can be mathematically removed from the Lorentz transformation to produce an alternate transformation that describes relativistic effects in the absence of differential simultaneity.  Differential simultaneity is shown to be required for Lorentz symmetry.  The alternate transformation is used as a contrast for the Lorentz transformation to assess whether relativistic experiments provide definitive evidence for differential simultaneity.  If the alternate transformation is compatible with an experiment, then the experiment does not provide definitive evidence for differential simultaneity because it can be explained in the absence of differential simultaneity.  Here, the alternate transformation is shown to be compatible with the broad array of current relativistic experiments.  This implies that current experiments do not provide definitive evidence for differential simultaneity.  Thus, an essential aspect of Lorentz symmetry currently lacks definitive experimental support.

\end{abstract}

\ccode{PACS numbers: 03.30.+p; 11.30.Cp}

%\tableofcontents

\section{Introduction}\label{section:1}
The Lorentz transformation (LT) forms the basis of special relativity (SR) and directly describes three relativistic effects: time dilation (TD); length contraction; and differential simultaneity (also known as the relativity of simultaneity) \cite{1}.  The simultaneity framework defines how time and space interact in response to motion.  Differential simultaneity involves time being offset with distance in ``moving'' frames (as viewed by a ``stationary'' observer) \cite{2}.

It is widely believed that all aspects of Lorentz invariance have been demonstrated with overwhelming experimental support \cite{3}.  Nevertheless, limited violations of Lorentz invariance are still being actively sought \cite{3}.  The significant interest in limited violations of Lorentz invariance arises from the presence of such violations in theories that propose to unify quantum effects and gravity, including string theory, non-commutative geometry, loop quantum gravity, and warped brane models \cite{4}.  To date, limited, context-dependent violations of Lorentz invariance have not been discovered. 

Tests of Lorentz invariance have addressed many possible limited violations of Lorentz invariance.  The standard model extension (SME) test theory considers over 80 different coefficients for violations of Lorentz invariance in the matter, photon, and gravity sectors \cite{5}.  However, none of these potential limited violations of Lorentz invariance check for the complete absence of differential simultaneity.  That is because the SME assumes observer Lorentz invariance (oLI), which implies that different observers view relativistic effects equivalently when rotated or in relative motion \cite{6}.  As described in Sec. \ref{section:2}, the complete absence of differential simultaneity causes different observers to view relativistic effects differently, thereby violating oLI.  Thus, the SME is incapable of assessing the complete absence of differential simultaneity. 

Two other relativistic test theories are more appropriate for addressing the presence or absence of differential simultaneity: the Mansouri and Sexl \cite{7, 8, 9} and the revised Robertson \cite{10} test theories.  Both test theories have a parameter that describes the global simultaneity framework, and whose value can be experimentally determined \cite{11, 12}.

There is a long-standing viewpoint that the LT is not applicable to rotational motion \cite{13}.  Despite this viewpoint, there is abundant evidence that relativistic effects occur in response to rotational motion.  Relativistic equations derived from the LT accurately describe rotational relativistic effects and are regularly cited to explain the relativistic effects, e.g. see Refs. 14--16.  Thus, kinematic transformations, such as the LT, can describe relativistic effects due to rotational motion.

It is often stated that Lorentz invariance is only present in the presence of gravity as local Lorentz invariance, which utilizes the LT framework to describe spacetime \cite{17}.  Local Lorentz invariance is proposed to occur in infinitesimal regions that approximate flat spacetime \cite{17}.  However, relativistic relations derived from the LT are routinely used to describe relativistic effects that occur in the presence of gravity at measurable (noninfinitesimal) distances.  Thus, kinematic transformations can be utilized to describe relativistic effects in the presence of gravity.

This study explores the extent that relativistic studies provide evidence for differential simultaneity.  The approach is to mathematically remove differential simultaneity from the LT to produce a transformation that is structurally equivalent to the LT but does not incorporate differential simultaneity.  The alternate transformation still describes the relativistic effects of length contraction and TD.  The alternate transformation is used as a contrast to the LT to determine if experiments provide definitive evidence for differential simultaneity.  If the alternate transformation is compatible with experimental data, then those experiments do not provide definitive evidence for differential simultaneity.  This study will consider all types of relativistic data, regardless of the type of motion or the presence of gravity.

Sections \ref{section:2} and \ref{section:3} describe the alternate transformation and its incorporation of preferred reference frames (PRFs).  Section \ref{section:8} provides a broad overview of how the alternate transformation is compatible with experimental tests of general relativity (GR) and quantum mechanics.  Section \ref{section:4} assesses the compatibility of the alternate transformation with specific types of relativistic experiments.  Section \ref{section:6} provides a discussion that describes approaches that can be used going forward to experimentally assess the simultaneity framework.

\section{Removing Differential Simultaneity from the LT}\label{section:2}
From the ``stationary''-frame perspective, the LT coordinate equations are:
\begin{equation}
t' = \frac{{t - \frac{{vx}}{{{c^2}}}}}{{\sqrt {1 - \frac{{{v^2}}}{{{c^2}}}} }},
\label{eq:1}
\end{equation}
\begin{equation}
x' = \frac{{x - vt}}{{\sqrt {1 - \frac{{{v^2}}}{{{c^2}}}} }},
\label{eq:2}
\end{equation}
\begin{equation}
y'{\rm{ }} = {\rm{ }}y,
\label{eq:3}
\end{equation}
\begin{equation}
z'{\rm{ }} = {\rm{ }}z,
\label{eq:4}
\end{equation}
where $t$ is time; $v$ is velocity; $x$, $y$, and $z$ are spatial coordinates; $c$ is the speed of light (SOL); and primed values represent the ``moving'' frame, while unprimed values represent the ``stationary'' frame.  

The LT equations directly describe three relativistic attributes: TD; length contraction; and differential simultaneity \cite{1}.  Differential simultaneity arises because the time coordinate equation (\ref{eq:1}) includes the distance term $x$.  This causes a ``stationary'' observer to view time in the ``moving'' frame to be a function of distance.  On a spacetime diagram, differential simultaneity causes the $x'$-axis to have a positive slope (Fig. 1a).  While spatially separated points on the $x'$-axis all have ``moving''-frame time $t'=0$, these points have different ``stationary'' times, which vary as a function of distance from the ``stationary'' observer.  This offsetting of time with distance generates different conclusions for what events are simultaneous between ``stationary'' and ``moving'' observers (Fig. 1a), hence the term ``differential simultaneity''.

\begin{figure}[ht]
\begin{center}
\includegraphics[width=11 cm]{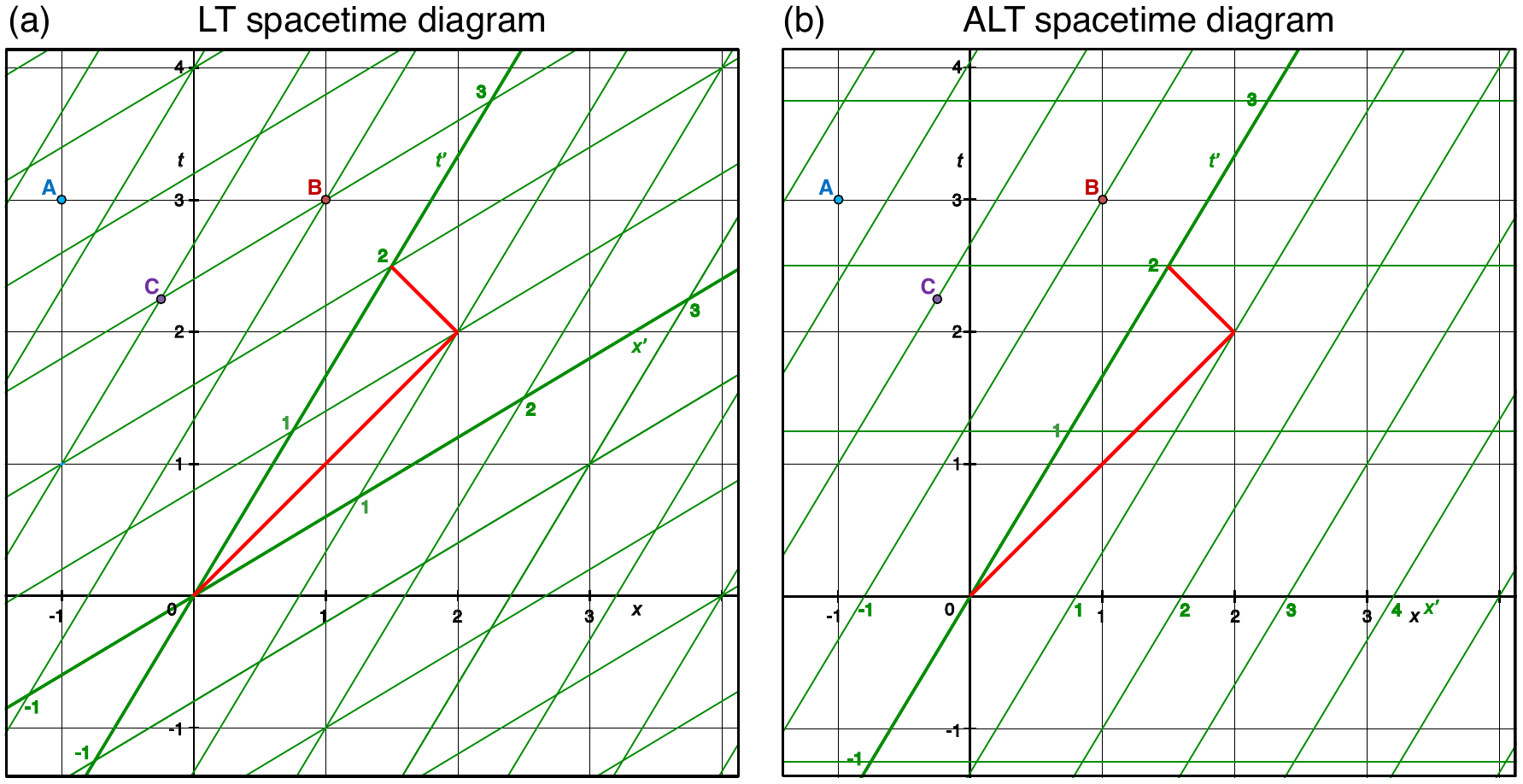}
\end{center}
\caption{Spacetime diagrams for the LT (a) and ALT (b).  Time is in seconds and distance is in light-seconds, $v = 0.6c$. ``Moving'' coordinate lines are green, and ``stationary'' coordinate lines are black.  The red line represents a light signal sent from $x'=0$ to $x'=1$ where it is reflected back to $x'=0$. (a) In the LT diagram, the one-way SOL in the ``moving'' frame is isotropic, i.e. 1 light-second in 1 s.  Events (A) and (B) are simultaneous for the ``stationary'' observer, while events (C) and (B) are simultaneous for the ``moving'' observer.  (b) In the ALT diagram, the one-way SOL is anisotropic, i.e. 1 light-second in 1.6 s forward and 0.4 s backward; but the two-way SOL is isotropic, i.e. 2 light-seconds in 2 s.  Events (A) and (B) are simultaneous for both stationary and moving observers.}
\label{figure1}
\end{figure}

As viewed by a ``stationary'' observer, objects in the ``moving'' frame are offset in time.  On a spacetime diagram, this manifests as ``moving'' objects being tilted in ``stationary''-frame time (Fig. 2a).  From the ``stationary'' perspective, ``moving''-frame time is offset by the term $-vl'/c\textsuperscript{2}$, where $l'$ is the distance between points in the ``moving'' frame (Fig. 2a) \cite{1}.  A ``stationary'' observer views the front of a ``moving'' object to have an earlier ``moving''-frame time than the back of the object.

\begin{figure}[ht]
\begin{center}
\includegraphics[width=11cm]{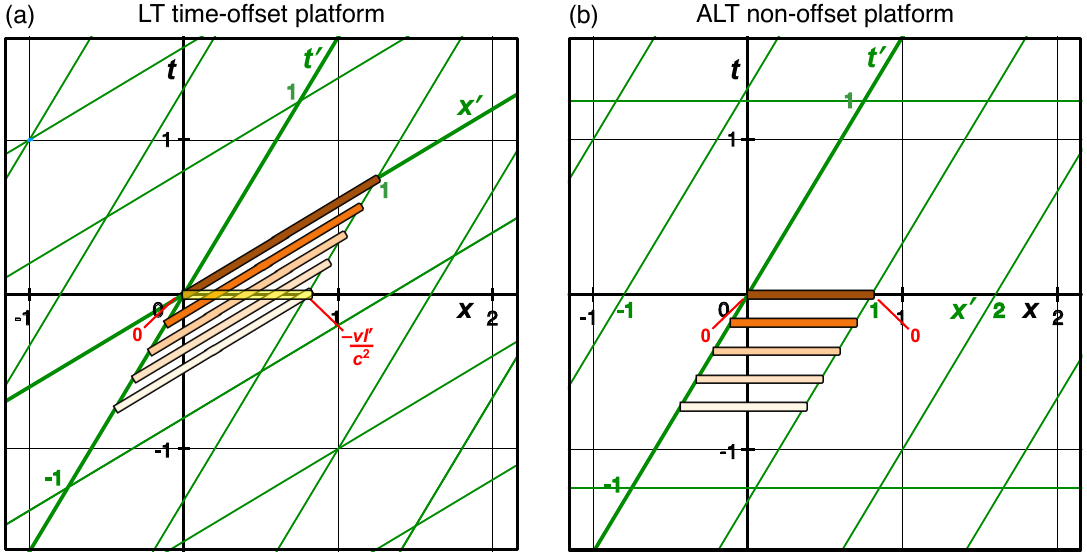}
\end{center}
\caption{``Moving''-frame time is offset with distance in the LT framework.  A ``moving'' platform is shown at five time points (lighter to darker brown) that span earlier-to-later ``moving''-frame times between $-0.6$ s to 0 s.  The convention of ``intelligent'' observers, who can take into account the transit times of light, is followed \cite{18}.  The ``moving'' platform contains synchronized clocks at its back and front.  A ``stationary'' observer at position 0,0 can observe the ``moving''-frame time on the clocks in the platform (shown in red lettering).  (a) In the LT diagram, the ``moving'' platform is offset in ``stationary'' time.  The 0,0 ``stationary'' observer views the platform at time 0 as the yellow, semi-transparent platform that is continuously shifted in time over its length.  The back of the platform is at $t' = 0$  while the front is at the earlier $t' = -vl'/c\textsuperscript{2}$ (here $-0.6$ s) \cite{1}.  (b) In the ALT diagram, the 0,0 stationary observer views the entire moving-frame platform to have $t' = 0$.}
\end{figure}

The offsetting of time with distance that generates differential simultaneity is described by the term $-vx'/c\textsuperscript{2}$ \cite{1}, which is called the relativistic time offset (RTO) \cite{12}.  Differential simultaneity can be removed from the LT $t'$ coordinate equation (\ref{eq:1}) by subtracting the RTO \cite{12}:  
\begin{equation}
{t'_{{\rm{LT - RTO}}}} = \frac{{t - \frac{{vx}}{{{c^2}}}}}{{\sqrt {1 - \frac{{{v^2}}}{{{c^2}}}} }} + \frac{{vx'}}{{{c^2}}} = t\sqrt {1 - \frac{{{v^2}}}{{{c^2}}}}.
\label{eq:20}
\end{equation}
The resulting $t'$ coordinate equation (\ref{eq:20}) does not have a distance term, and so does not offset time with distance to generate differential simultaneity.

The simultaneity framework only affects the $t'$ coordinate equation, and the other transformation equations are not altered when differential simultaneity is removed.  Thus, the full transformation without differential simultaneity is
\begin{equation}
t' = t\sqrt {1 - \frac{{{v^2}}}{{{c^2}}}},
\label{eq:12}
\end{equation}
\begin{equation}
x' = \frac{{x - vt}}{{\sqrt {1 - \frac{{{v^2}}}{{{c^2}}}} }},
\label{eq:27}
\end{equation}
\begin{equation}
y'{\rm{ }} = {\rm{ }}y,
\label{eq:28}
\end{equation}
\begin{equation}
z'{\rm{ }} = {\rm{ }}z.
\label{eq:29}
\end{equation}
Plotting these coordinate equations on a spacetime diagram produces a horizontal $x'$-axis, reflecting that time is not offset with distance, as expected for the absence of differential simultaneity (Fig. 1b).  Accordingly, moving objects are not offset in time over their length when viewed from the stationary frame (Fig. 2b).  Notably, the alternate transformation still describes the relativistic effects of TD and length contraction in the moving frame, as shown by the altered spacing of units on the moving-frame $x'$- and $t'$-axes relative to the stationary-frame (Fig. 1b).  

The transformation without differential simultaneity is referred to as the absolute LT (ALT) because the absence of differential simultaneity is referred to as absolute simultaneity.  

Since the LT and ALT only differ structurally in whether differential simultaneity is present or absent, a comparison of the two transformations can determine the importance of differential simultaneity for Lorentz symmetry.  To assess Lorentz symmetry, it is important to consider how relativistic effects are observed from the ``moving''-frame perspective.  The ``moving''-frame perspective is described by the converse ``moving''-frame coordinate transformation equations.  The LT ``moving''-frame equations are obtained by rearrangement of the LT ``stationary''-frame equations (\ref{eq:1})--(\ref{eq:4}) to give
\begin{equation}
t = \frac{{t' + \frac{{vx'}}{{{c^2}}}}}{{\sqrt {1 - \frac{{{v^2}}}{{{c^2}}}} }},
\label{eq:5}
\end{equation}
\begin{equation}
x = \frac{{x' + vt'}}{{\sqrt {1 - \frac{{{v^2}}}{{{c^2}}}} }},
\label{eq:6}
\end{equation}
\begin{equation}
y = y',
\label{eq:7}
\end{equation}
\begin{equation}
z = z'.
\label{eq:8}
\end{equation}
These equations are the symmetrical equivalents of the LT ``stationary''-frame equations  (\ref{eq:1})--(\ref{eq:4}) when considering that $v$ is of opposite sign.  Lorentz symmetry is reflected in that both ``stationary'' and ``moving'' LT observers use the same coordinate equations to describe the other reference frame.  

In contrast, rearranging the ALT stationary-frame equations (\ref{eq:12})--(\ref{eq:29}) generates different moving-frame equations \cite{19}:
\begin{equation}
t = \frac{{t'}}{{\sqrt {1 - \frac{{{v^2}}}{{{c^2}}}} }},
\label{eq:13}
\end{equation}
\begin{equation}
x = x'\sqrt {1 - \frac{{{v^2}}}{{{c^2}}}}  + \frac{{vt'}}{{\sqrt {1 - \frac{{{v^2}}}{{{c^2}}}} }},
\label{eq:14}
\end{equation}
\begin{equation}
y = y',
\label{eq:15}
\end{equation}
\begin{equation}
z = z'.
\label{eq:16}
\end{equation}
The differences between the ALT moving-frame equations (\ref{eq:13})--(\ref{eq:16}) and stationary-frame equations (\ref{eq:12})--(\ref{eq:29}) indicates that moving and stationary observers use different coordinate equations to describe the other reference frame.  This implies the absence of Lorentz symmetry.

Relativistic effects are described by relations.  From the ``stationary''-frame perspective, the LT and ALT share the relations for TD, length contraction, and increases in relativistic mass/energy \cite{7, 20}:
\begin{equation}
\Delta t' = \Delta t\sqrt {1 - \frac{{{v^2}}}{{{c^2}}}} ,
\label{eq:9}
\end{equation}
\begin{equation}
\Delta l' = \Delta l\sqrt {1 - \frac{{{v^2}}}{{{c^2}}}} ,
\label{eq:10}
\end{equation}
\begin{equation}
\Delta m' = \frac{{\Delta m}}{{\sqrt {1 - \frac{{{v^2}}}{{{c^2}}}} }},
\label{eq:11}
\end{equation}
where $l$ is length, and $m$ is mass.  

The LT ``moving''-frame relativistic relations are the symmetrical equivalent of the LT/ALT ``stationary''-frame relations (\ref{eq:9})--(\ref{eq:11}), differing only in having the primed and unprimed terms switched:
\begin{equation}
\Delta t = \Delta t'\sqrt {1 - \frac{{{v^2}}}{{{c^2}}}} ,
\label{eq:30}
\end{equation}
\begin{equation}
\Delta l = \Delta l'\sqrt {1 - \frac{{{v^2}}}{{{c^2}}}} ,
\label{eq:31}
\end{equation}
\begin{equation}
\Delta m = \frac{{\Delta m'}}{{\sqrt {1 - \frac{{{v^2}}}{{{c^2}}}} }}.
\label{eq:32}
\end{equation}
Due to this symmetry, two LT observers in different inertial reference frames (IRFs) observe equivalent, reciprocal relativistic effects when viewing each other.  This exemplifies Lorentz symmetry.  The use of quotation marks for ``stationary'' and ``moving'' denotes the interchangeability of these designations in the LT framework.

In contrast, the ALT moving-frame relations are the inverse of the ALT/LT stationary-frame relations (\ref{eq:9})--(\ref{eq:11}):
\begin{equation}
\Delta t = \frac{{\Delta t'}}{{\sqrt {1 - \frac{{{v^2}}}{{{c^2}}}} }},
\label{eq:17}
\end{equation}
\begin{equation}
\Delta l = \frac{{\Delta l'}}{{\sqrt {1 - \frac{{{v^2}}}{{{c^2}}}} }},
\label{eq:18}
\end{equation}
\begin{equation}
\Delta m = \Delta m'\sqrt {1 - \frac{{{v^2}}}{{{c^2}}}} .
\label{eq:19}
\end{equation}
The ALT moving-frame relations describe inverse relativistic effects when a moving-frame observer views a stationary frame, namely: time contraction (i.e. a faster passage of time); length expansion; and decreased relativistic mass/energy.  Since the ALT moving and stationary frames can be experimentally distinguished based on the directionality of relativistic effects, the two frames are not interchangeable, and thus are written without quotation marks.  This directionality is incompatible with Lorentz symmetry.  Overall, the comparisons in this section show that removing differential simultaneity from the LT abolishes Lorentz symmetry.

\subsection{Simultaneity does not alter ``stationary'' frame kinematics}\label{section:2.1}
The ALT transformation has been previously studied.  ALT was first described in 1938 \cite{21}, and has been independently discovered at least five additional times \cite{22, 7, 23, 24, 19}.  The transformation was given the name ALT in 1958 \cite{22}.  A number of studies have described ALT kinematics.  These show that ALT shares multiple relativistic kinematics with the LT from the stationary perspective, including: TD \cite{22, 7, 19}; length contraction \cite{22, 7, 19}; relativistic mass/energy \cite{20}; relativistic Doppler shift \cite{25, 26}; relativistic stellar aberration angle \cite{25, 26}; the one-way SOL being independent of the motion of its source \cite{22}; the isotropic two-way SOL \cite{22}; and the one-way SOL $c$ in the ``stationary'' frame \cite{7}.  

When measurements are made from the ``stationary'' perspective, ALT has been determined to be indistinguishable from the LT \cite{7, 8, 9}.  Thus, relativistic effects from the ``stationary'' perspective are not affected by the presence or absence of differential simultaneity.

Analyses of all potential LT-related transformations found that only the LT and ALT correctly predict the results of three classic tests of relativity \cite{27, 28}.  This suggests that the LT and ALT are the only potentially viable Lorentz-like transformations.

\section{The PRF Framework for ALT}\label{section:3}
The ALT framework includes preferred reference frames (PRFs) \cite{22, 7, 19}.  In a PRF framework, relativistic velocities are calculated relative to a PRF \cite{22, 7, 19}.  ALT predicts directional relativistic effects between observers based on velocities relative to the PRF.  To understand this, consider two inertial observers, with observer 2 moving faster relative to a PRF, and observer 1 moving slower.  Since observer 2’s motion is greater relative to the PRF, observer 2 experiences more TD than observer 1.  Observer 1 views that (relative to them) observer 2 experiences TD.  Conversely, observer 2 views that observer 1 experiences time contraction.  Thus, both observers agree on the directionality of the relativistic effects.  

Historically, ALT was proposed to have a single, absolute PRF, which was often assumed to be the cosmic microwave background (CMB) \cite{7, 27}.  However, an external PRF such as the CMB would mean that the Earth's orbital and rotational motions ($\sim30$ km/s and $\sim0.4$ km/s, respectively) would alter velocities relative to the external PRF (Fig. 3a).  These altered velocities would make ALT predictions incompatible with high-resolution experiments that show that potential external PRFs do not impact Lorentz symmetry or relativistic effects on the Earth's surface, e.g. see Refs. 29 and 30.

\begin{figure}[ht]
\begin{center}
\includegraphics[width=10cm]{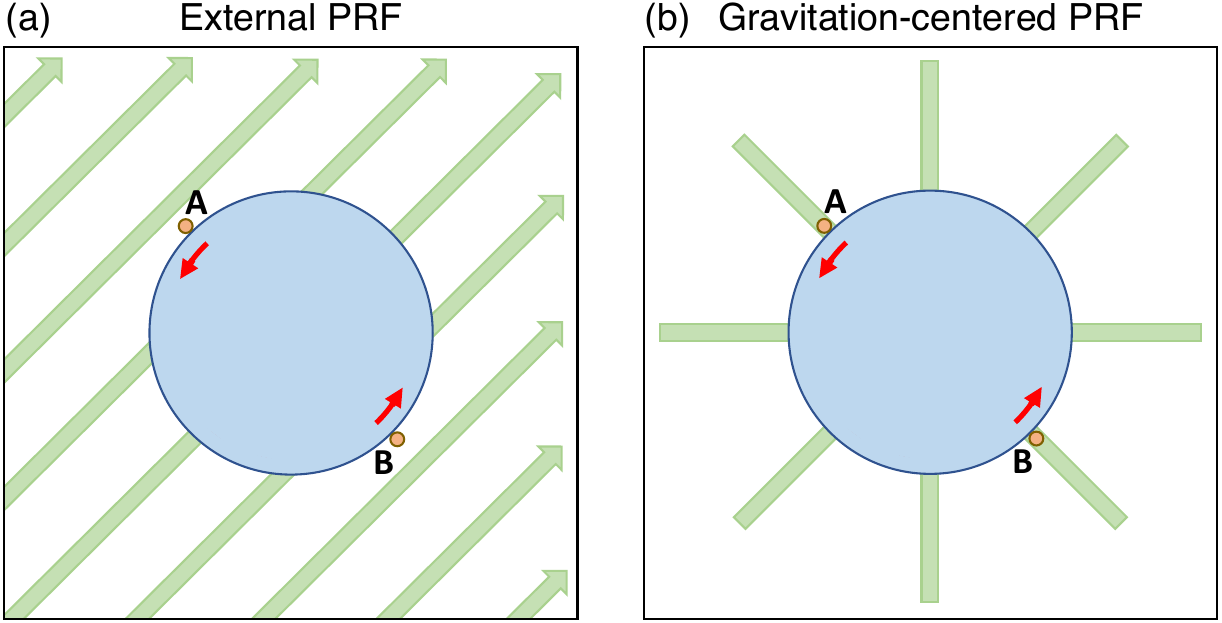}
\end{center}
\caption{External and gravitation-centered PRFs.  A planet is viewed through its axis of rotation.  (a) A diagram of an external PRF that has a velocity (denoted by green arrows) relative to the rotating planet.  Point (B) on the surface is moving in the same direction as the PRF, and has a smaller velocity relative to the PRF.  Point (A) is moving in the opposite direction as the PRF, and has a larger velocity relative to the PRF.  (b) In a nonrotating, gravitation-centered PRF, points (A) and (B) have the same velocity relative to the PRF.}
\label{Fig3}
\end{figure}

In 2014, it was proposed that PRFs for ALT are locally associated with centers of gravitational mass \cite{31} (Fig. 3b).  This PRF framework allows ALT to be compatible with diverse relativistic experiments \cite{11, 32}.  The combination of ALT with gravitation-centered PRFs is referred to as absolute simultaneity theory (AST) \cite{31}.  

The local PRF for AST is the nonrotating Earth-centered inertial (ECI) reference frame, which is centered on the Earth's gravitational core.  There is experimental evidence that the ECI functions as a PRF, relative to which velocities are calculated and directionality is determined.  Notably, the one-way SOL has been demonstrated to be isotropic in the ECI, but anisotropic in the rotating Earth-centered Earth-fixed (ECEF) reference frame \cite{33, 34, 35}.  Additionally, TD has been shown to be directional, with the extent of TD determined based on velocity relative to the ECI \cite{14, 15}.

AST predicts that when events are within the solar system but distant from planets, then the sun-centered heliocentric inertial (HCI) reference frame is the relevant PRF.  This prediction is supported by the observation that one-way light signals traversing the solar system are isotropic from the HCI perspective but not from the ECI perspective \cite{36}.

\subsection{The HCI does not function as a PRF near the Earth } \label{section:3.1}
On or near the Earth, the local PRF in the AST framework is the ECI.  This contrasts with published Lorentz invariance experiments that use the HCI as the reference frame to analyze experiments on the Earth \cite{37}.  Notably, analyses of velocity-dependent relativistic effects on or near the Earth are consistent with the ECI acting as a PRF, but not with the HCI.  To illustrate this, consider an idealized Hafele and Keating experiment \cite{14, 15}.  Here, the flights of the two planes are eastward and westward at the equator, and the velocities of the planes in both directions are 464 m s\textsuperscript{-1}, which is equivalent to the Earth's orbital velocity at the equator.  The westward plane has a velocity of 0 m s\textsuperscript{-1} relative to the ECI (because it is moving at 464 m s\textsuperscript{-1} westward relative to the Earth's surface, while the Earth's surface is rotating 464 m s\textsuperscript{-1} eastward, and so the plane stays in the same position relative to the ECI).  The eastward plane has a velocity of 928 m s\textsuperscript{-1} relative to the ECI  (because the plane's velocity relative to the Earth's surface is 464 m s\textsuperscript{-1} eastward and this is added to the Earth's surface velocity of 464 m s\textsuperscript{-1} eastward relative to the ECI).  The two planes take off at 6 PM local time (when the starting point is in line with the Earth's orbital trajectory).  The planes meet again 12 h later, and their clocks are directly compared.  

In the calculations below, the effect of reduced gravitational TD (because the planes are flying at altitudes above the Earth's surface) are not included.  Gravitational TD affects both planes equally, and the effects are well known and so can be factored into the calculation \cite{14, 15}.  Their inclusion here would just complicate the discussion without bringing new insights.

Using the shared ALT/LT TD equation (\ref{eq:9}) with velocities calculated relative to the ECI, the eastward clock is predicted to be 207 ns slower than the westward clock due to TD.  If the experiment is instead analyzed using velocities relative to the HCI, the westward clock has the same velocity as the Earth's orbital velocity, 29,780 m s\textsuperscript{-1}.  The eastward clock has a higher velocity of 30,708 m s\textsuperscript{-1} due to the extra movement around the Earth's circumference.  When using the HCI velocities, the eastward clock is predicted to be 13,459 ns slower than the westward clock when the two planes meet.  Thus, using the HCI velocities generates a 65-fold larger timing difference than the use of ECI velocities.  Hafele and Keating's experiment predicted a timing difference of 315 ns using ECI velocities and they observed a difference of 332 ns, i.e. an observed to predicted ratio of 1.05 \cite{14, 15}.  Their experiment had sufficient resolution to have detected a 65-fold larger timing difference (as would have occurred if the HCI was the relevant frame).  Thus, the Hafele and Keating experiment demonstrates that the ECI acts as the PRF near the Earth, and rules out the HCI as the PRF near the Earth, both of which are predicted by AST.

\subsection{Experimental limits on how gravitation could affect AST}\label{section:3.2}
The theoretical basis for why gravitational centers would act as PRFs in the AST framework is not known.  However, a link between gravitational centers and PRFs is not unprecedented.  In GR, gravitational centers determine the directionality of relativistic effects.  For example, all observers agree that observers closer to a gravitational center experience greater TD than observers farther from the gravitational center.  There are, however, notable differences in how gravity impacts relativistic effects for GR and AST.  With GR, differences in gravitational force determine the extent of relativistic effects.  The potential impact of local variations in gravity on AST predictions is discussed here. 

The accuracy of the global positioning system (GPS) requires corrections to the timing of clocks in GPS satellites to account for relativistic time-dilation effects.  Relativistic timing differences in GPS clocks (relative to the Earth's surface) arise from increased TD from orbital velocity (which \textit{decreases} GPS clock timing by 7210 ns per day) and a decrease in gravitational TD (which \textit{increases} GPS clock timing by 45,650 ns per day) \cite{38}.  The two relativistic effects are additive and together produce an overall increase in timing of 38,440 ns per day relative to the Earth's surface \cite{38}.  While the change in gravitational TD has the larger effect, in the absence of the motion-based TD correction, GPS would have a localization error of 2.16 km per day \cite{38}.  GPS atomic clocks are set at a lower frequency prior to launch to exactly counter the additive effects of the satellite's motion coupled with the effects of the lower gravitational force \cite{39, 40}.  

In contrast to the 38,440 ns per day alteration of clock timing to address the relativistic effects, the average daily error in GPS clock timing is only 13.7 ns per day \cite{41}.  The daily accuracy of GPS clocks is thus a 0.0019 ratio relative to the correction for just the motion-based TD (i.e. 13.7 ns/7210 ns).  Given that there are many factors that can affect GPS clock accuracy (including frequency drift, differences in gravitational potential, and altered satellite position \cite{42, 43}), the accuracy of the frequency correction for the motion-based TD is likely to be substantially better than the 0.0019 ratio.  If the extent of the motion-based TD was measurably different, then it would systemically affect the operation of all GPS satellites, and the frequency programmed into the clocks would need to be altered to reflect that difference. 

The analysis of GPS satellite timing thus suggests that the application of the ALT/LT shared TD equation (\ref{eq:30}) at the altitude of GPS satellites (20,183 km \cite{38, 40}) is accurate to at least a resolution of 0.0019, but probably significantly lower.  This suggests that there is not an obvious change in the motion-based TD in response to the reduction in the force of gravity at the altitude of GPS satellites.  Thus, experiments carried out on the Earth's surface or in Earth orbit can be analyzed with AST without the necessity to consider the effects of gravity.

\subsection{Simultaneity alters the one-way SOL and Sagnac effect}\label{section:2.2}
In a PRF framework, the one-way SOL is isotropic for observers in the PRF but is anisotropic for non-PRF observers \cite{7}.  Thus, with AST, observers in moving or rotating frames (that are not PRFs) observe anisotropic one-way light speeds.  In a rotating frame, the anisotropic one-way SOL produces the Sagnac effect.  Here the one-way SOL, the two-way SOL, and the Sagnac effect are derived for ALT and the LT.

The one-way speeds of light associated with each transformation can be derived directly from their distance ($x'$) and time ($t'$) coordinate transformation equations, noting that $\Delta$$x$/$\Delta$$t$ = $c$, where $\Delta$$x$ is the ``stationary''-frame distance that is traversed by a light signal, and $\Delta$$t$ is the ``stationary''-frame time for the light signal to traverse that distance.  

The LT one-way SOL based on its coordinate equations (\ref{eq:1}) and (\ref{eq:2}) is isotropic $c$ (in the forward and reverse directions):
\begin{equation}
c{'_{{\rm{LT}}}} = \frac{{\Delta x'}}{{\Delta t'}} = \frac{{\frac{{\Delta x - v\Delta t}}{{\sqrt {1 - \frac{{{v^2}}}{{{c^2}}}} }}}}{{\frac{{\Delta t - \frac{{v\Delta x}}{{{c^2}}}}}{{\sqrt {1 - \frac{{{v^2}}}{{{c^2}}}} }}}} = c;c.
\label{eq:41}
\end{equation}

The ALT one-way SOL based on its coordinate equations (\ref{eq:12}) and (\ref{eq:27}) (in the forward and reverse directions) is
\begin{equation}
{c'_{{\rm{ALT}}}} = \frac{{\Delta x'}}{{\Delta t'}} = \frac{{\frac{{\Delta x - v\Delta t}}{{\sqrt {1 - \frac{{{v^2}}}{{{c^2}}}} }}}}{{\Delta t\sqrt {1 - \frac{{{v^2}}}{{{c^2}}}} }} = \frac{c}{{1 + \frac{v}{c}}};\frac{c}{{1 - \frac{v}{c}}}.
\label{eq:42}
\end{equation}

The two-way speeds of light associated with each transformation can be derived by substituting their one-way speeds of light into the basic two-way SOL equation:
\begin{equation}
c{'_{{\rm{two-way}}}} = {\rm{ }}\frac{{l' + l'}}{{t{'_{{\rm{forward}}}} + t{'_{{\rm{backward}}}}}} = {\rm{ }}\frac{{l' + l'}}{{\frac{{l'}}{{c{'_{{\rm{forward}}}}}} + \frac{{l'}}{{c{'_{{\rm{backward}}}}}}}},
\label{eq:43}
\end{equation}
where $l'$ is the ``moving''-frame distance that light traverses in each direction, and $t'$ = $l'$/$c'$.

The LT two-way SOL is isotropic $c$:
\begin{equation}
c{'_{{\rm{two-wayLT}}}} = \frac{{l' + l'}}{{\frac{{l'}}{c} + \frac{{l'}}{c}}} = c.
\label{eq:44}
\end{equation}

The ALT two-way SOL is also isotropic $c$, as previously reported \cite{22, 44}:
\begin{equation}
c{'_{{\rm{two-wayALT}}}} = \frac{{l' + l'}}{{\frac{{l'}}{{\left( {\frac{c}{{1 + \frac{v}{c}}}} \right)}} + \frac{{l'}}{{\left( {\frac{c}{{1 - \frac{v}{c}}}} \right)}}}} = c.
\label{eq:45}
\end{equation}

The Sagnac effect can be derived from the rotating-frame one-way SOL.  The rotational form of the LT is the Franklin transformation \cite{45}, and the rotational form of ALT has been previously described \cite{44}.  The one-way speeds of light for each rotational transformation are equivalent to that of the linear except for the use of polar coordinates, where peripheral velocity is $\omega r$ and distance is $d\theta r$ \cite{44, 32}.  The rotating-frame one-way speeds of light are derived using the coordinate transformation equations, as above.  

The LT/Franklin one-way SOL in a rotating frame (in the co-rotating and counter-rotating directions) is
\begin{equation}
c{'_{{\rm{LT}}}} = \frac{{d\theta 'r'}}{{dt'}} = \frac{{\frac{{d\theta r - v(r)dt}}{{\sqrt {1 - \frac{{v{{(r)}^2}}}{{{c^2}}}} }}}}{{\frac{{dt - \frac{{v(r)d\theta r}}{{{c^2}}}}}{{\sqrt {1 - \frac{{v{{(r)}^2}}}{{{c^2}}}} }}}} = c;c.
\label{eq:46}
\end{equation}

The ALT one-way SOL in a rotating frame is
\begin{equation}
c{'_{{\rm{ALT}}}} = \frac{{d\theta 'r'}}{{dt'}} = \frac{{\frac{{d\theta r - \omega rdt}}{{\sqrt {1 - \frac{{{\omega ^2}{r^2}}}{{{c^2}}}} }}}}{{dt\sqrt {1 - \frac{{{\omega ^2}{r^2}}}{{{c^2}}}} }} = \frac{c}{{1 + \frac{{\omega r}}{c}}};\frac{c}{{1 - \frac{{\omega r}}{c}}}.
\label{eq:47}
\end{equation}

The Sagnac effect is the time for light to go around a disk in the co-rotating direction minus the time in the counter-rotating direction:
\begin{equation}
{\rm{Sagnac}} = t{'_{{\rm{co - rotating}}}} - t{'_{{\rm{counter - rotating}}}} = \frac{{l'}}{{c{'_{{\rm{co - rotating}}}}}} - \frac{{l'}}{{c{'_{{\rm{counter - rotating}}}}}}.
\label{eq:48}
\end{equation}

Substituting the LT one-way SOL (\ref{eq:46}) into Eq. \ref{eq:48}, with a distance of $2\pi r$, gives a null Sagnac effect:
\begin{equation}
{\rm{Sagna}}{{\rm{c}}_{{\rm{LT}}}} = \frac{{2\pi r}}{c} - \frac{{2\pi r}}{c} = 0.
\label{eq:49}
\end{equation}

Substituting the ALT one-way speeds of light (\ref{eq:47}) into equation \ref{eq:48}, with a distance of $2\pi r$, gives the conventional Sagnac effect equation:
\begin{equation}
{\rm{Sa}}{{\rm{g}}_{{\rm{2AS}}}} = \frac{{2\pi r}}{{\left( {\frac{c}{{1 + \frac{{\omega r}}{c}}}} \right)}} - \frac{{2\pi r}}{{\left( {\frac{c}{{1 - \frac{{\omega r}}{c}}}} \right)}} = \frac{{4\pi \omega {r^2}}}{{{c^2}}}.
\label{eq:55}
\end{equation}

Therefore, ALT and the LT are distinguished by the one-way SOL and the Sagnac effect, but not by the two-way SOL.

\section{The Simultaneity Framework and GR and Quantum Mechanics}\label{section:8}
Before the determination of the compatibility of AST with specific categories of relativistic experiments, it is worthwhile to have a general discussion on the absence of differential simultaneity and the experimental support for GR and quantum mechanics.

\subsection{The simultaneity framework and GR at the global level }\label{section:8.1}
Removing differential simultaneity from the LT abolishes its group structure.  GR at the global level describes gravitation using the field equations, and at this level, GR also does not have a group structure \cite{46}.  Thus, the absence of a group structure in ALT is shared by GR at the global level.  

Removing differential simultaneity from the LT abolishes the relativity principle.  Notably, it is the predominant view among relativists that GR at the global level and its principle of general covariance also do not encompass the relativity principle \cite{47, 48, 49}.  Manifestly, GR describes directional relativistic effects (rather than reciprocal effects), and these directional effects are at odds with the relativity principle.  For example, consider two observers at different distances from a gravitational center.  The observer nearer to the gravitational center experiences more TD, and the observer farther from the center experiences less TD.  Thus, relative to each other, the nearer observer experiences TD, and the farther observer experiences time contraction, with both observers agreeing on the directionality.  This directionality of relativistic effects is incompatible with the relativity principle, for which relativistic effects should be symmetrical.  Thus, both ALT and GR at the global level do not embody the relativity principle. 

\subsection{ ALT is a generally covariant solution for GR }\label{section:8.2}
It is proposed that GR exhibits Lorentz symmetry at infinitesimal local regions that are so small that they are not affected by gravity \cite{17}.  In the absence of curvature, the metric tensor of GR, $g$\textsubscript{$\mu v$}, is proposed to be equivalent to the metric tensor of SR, $\eta$\textsubscript{$\mu v$}.  Thus, in the absence of spacetime curvature (i.e. in infinitesimal regions), LT spacetime is considered to be the descriptor for GR \cite{50}.  An analysis has shown that $\eta$\textsubscript{$\mu v$} is the sole solution to GR in the absence of curvature, but only when including the stipulation that all uniformly translating frames are equivalent \cite{22}.  In the absence of this stipulation, the ALT metric tensor is also a generally covariant solution for GR in the absence of spacetime curvature \cite{22}.  The ALT metric tensor has been generalized to describe any direction of motion, not just motion in the $x$-direction \cite{22}.  ALT has also been described in a generally covariant manner with coordinate-independent language \cite{22}.

In the absence of curvature, GR spacetime from the ``stationary'' perspective is described by the Minkowski metric line element:
\begin{equation}
d{s^2} = {c^2}d{t^2} - d{x^2} - d{y^2} - d{z^2}.
\label{eq:50}
\end{equation}

Notably, the Minkowski metric is the line element for both ALT and the LT from the ``stationary'' perspective \cite{12}.  This can be mathematically illustrated.  The LT line element from the ``moving'' perspective is:
\begin{equation}
ds{'^2} = {c^2}dt{'^2} - dx{'^2} - dy{'^2} - dz{'^2}.
\label{eq:51}
\end{equation}
The ALT line element from the moving perspective is \cite{22}:
\begin{equation}
ds{'^2} = {c^2}dt{'^2} - 2vdx'dt' - \left( {1 - \frac{{{v^2}}}{{{c^2}}}} \right)dx{'^2} - dy{'^2} - dz{'^2}.
\label{eq:52}
\end{equation}
Substituting the values from the LT transformation equations (\ref{eq:1})--(\ref{eq:4}) into the ``moving''-frame line element equation (\ref{eq:51}), or substituting the values from the ALT transformation equations (\ref{eq:12})--(\ref{eq:29}) into the moving-frame line element equation (\ref{eq:52}), both produce the ``stationary''-frame Minkowski line element:
\begin{equation}
d{s^2} = {c^2}d{t^2} - d{x^2} - d{y^2} - d{z^2}.
\label{eq:53}
\end{equation}
Thus, both the LT and ALT are compatible with the ``stationary'' frame description of spacetime for GR in the absence of curvature.  

\subsection{The equivalence principle and differential simultaneity}\label{section:8.3}
The equivalence principle of GR has been described to encompass three components: i) the weak equivalence principle that states that mass is proportional to weight, and two objects fall with the same acceleration in a gravitational field regardless of their mass; ii) the outcome of any local nongravitational experiment is independent of the velocity of a freely falling apparatus; and iii) the outcome of any local nongravitational experiment is independent of where and when it is performed \cite{51, 52}.  Here, ``local nongravitational experiment'' indicates an experiment carried out in a freely falling laboratory/apparatus \cite{52}.  The second component of the equivalence principle implies that an observer in a free-fall reference frame views all relativistic effects as if they are ``at rest''.  

As described in Sec. \ref{section:2}, ALT and the LT can be distinguished only by observations within ``moving'' frames, but not by ``stationary''-frame observations \cite{7, 8, 9}.  Further, only certain relativistic effects are altered by the simultaneity framework, these include the one-way SOL and the directionality of velocity-dependent relativistic effects.  AST predicts violations of the second component of the equivalence principle for free-fall frames that are not AST PRFs.  In particular, AST predicts light speed anisotropy and directional relativistic effects within a free-fall apparatus that is in motion relative to a PRF.  

Unfortunately, the specific experimental tests of the equivalence principle that could distinguish between AST and SR have not been carried out.  Rather, tests of the equivalence principle have focused on tests of the gravitational predictions of GR.  These include searches for violations of the weak equivalence principle and the universality of free fall, variations in the locally measured value of the gravitational constant or its alteration over time, the universality of gravitational redshift, and precision measurements of gravity \cite{52}.  GR does not incorporate differential simultaneity in its field equations, and so tests of the gravitational predictions of GR do not provide evidence for differential simultaneity.  Further, neither the LT nor ALT encompass gravity (instead describing relativistic effects in response to motion), and so the two transformations are not distinguished by analyses of gravity.

\subsection{ Quantum mechanics and differential simultaneity}\label{section:8.4}
SR has been incorporated into quantum mechanics from early in its development \cite{53}.  Quantum mechanics is a very successful theory, but the question here is whether any experimental tests of quantum mechanics have provided definitive evidence for differential simultaneity.  As described in Sec. \ref{section:2}, experimental evidence for differential simultaneity requires observations from within ``moving'' or rotating frames that probe velocity-dependent relativistic effects.  Experimental analyses of subatomic particles utilize the ``stationary'' frame perspective rather than making observations from the perspective of the particles themselves.  Since the LT and ALT share the same predictions for relativistic effects from the ``stationary'' perspective, these experiments do not provide insights into the simultaneity framework.  In the next section, relativistic experiments with subatomic particles are analyzed (along with other types of relativistic experiments) to determine if they provide definitive evidence for differential simultaneity.

There is a growing consensus that differential simultaneity is incompatible with a fundamental aspect of quantum mechanics: the transmission of information through quantum entanglement \cite{54, 55, 56}.  The offsetting of time with distance that gives rise to differential simultaneity causes space-like separated events (which cannot be connected by light signals) to not have a defined temporal order.  Therefore, which of two events occurs first can differ for different observers.  This reversal of temporal order is fundamentally incompatible with the Neumann-Dirac collapse formulation of quantum mechanics \cite{54, 55, 56}.  

\section{Compatibility of Relativistic Experiments with AST}\label{section:4}
In this section, a broad range of relativistic experiments are assessed for compatibility with AST.  The different types of relativistic experiments can be placed into six categories based on the reason for compatibility with AST.  Tables 1--6 provide a more extensive list of relativistic experimental types for each category that includes a representative reference, the reason for compatibility with AST, and information on whether the data was obtained from a rotating frame.

\subsection{Category 1: Irrelevant hypothetical Lorentz invariance violations}\label{section:4.1}
The experiments in category 1 test for potential Lorentz invariance violations (LIVs) that are not predicted by AST or SR, and are often not within their purview.  Since the potential LIVs are not predicted by AST or SR, the failure to observe the LIVs in these experiments does not impact the validity of AST or SR.  The number of potential LIVs is extensive, as documented in the SME \cite{5}.  Many of these potential LIVs are not related to relativistic effects in response to motion, which is the focus of AST and SR.  

\begin{table}[ht]
\footnotesize
\begin{flushleft}
\textbf{Table 1. Category 1: Hypothetical LIVs that are not relevant for AST or SR.}
\end{flushleft}
\label{Table1}
\rowcolors{1}{}{lightgray}
\begin{tabular}{lll}%%%The number of columns has to be defined here
\hline 
Experiment type &Rotation &AST compatibility \\
\hline
\makecell[tl]{Absorption of astrophysical gamma rays\\is not increased by hypothetical above-\\maximum velocities of charged particles \cite{57}}{\hspace{10.2mm}} & None & \makecell[tl]{Particles not faster than\\SOL within their PRF \cite{7}} {\hspace{5mm}}\\

\makecell[tl]{Photon energy does not alter\\one-way SOL \cite{58}} & None & \makecell[tl]{SOL is constant within its\\PRF \cite{7, 22}} \\

\makecell[tl]{High-energy photons do not decay or split\\as predicted for superluminal SOL \cite{59}} {\hspace{9mm}} & None & \makecell[tl]{SOL is constant within its\\PRF \cite{7, 22}} \\

\makecell[tl]{High-energy neutrinos do not exhibit\\anomalous flavor-changing effects \cite{60}} & None & \makecell[tl]{AST does not predict\\neutrino flavor changes} \\

\makecell[tl]{Very long baseline interferometry probing\\gravitational deflection of radio waves\\from solar system bodies \cite{61}} {\hspace{9.7mm}} & \makecell[tl]{Solar system\\rotation}{\hspace{3mm}} & \makecell[tl]{AST does not encompass\\gravitation (AST-DNEG)} \\

\makecell[tl]{Planetary ephemerides to analyze gravity \cite{62}} & \makecell[tl]{Solar system\\rotation} & AST-DNEG\\

Atom interferometry to analyze gravity \cite{63} & Earth rotation & AST-DNEG\\

Gravity Probe B gyroscopic tests of gravity \cite{64} & Earth rotation & AST-DNEG\\

Orbital dynamics of binary pulsars \cite{65} & Pulsar orbits & AST-DNEG\\

\makecell[tl]{Lunar laser ranging to analyze gravity \cite{66}} & None & AST-DNEG\\

Speed of gravitational waves vs. photons \cite{67} & None & AST-DNEG \\
\vspace*{-10pt}
\\\hline
\end{tabular}
\end{table}%%%End of the table

LIVs that are not within the purview of AST or SR include gravitational LIVs \cite{68} (Table 1).  These experiments are not relevant because both AST and SR do not encompass gravity, and GR at the global level does not incorporate differential simultaneity.  Therefore, experimental support for GR predictions on gravitation do not provide evidence for differential simultaneity. 

Given the broad scope of potential LIVs that are encompassed by category 1, and their nonapplicability to AST and SR, Table 1 does not list all category 1 experiments, but only those for gravitational experiments or those that impact a basic precept of AST or SR, such as the SOL.  Potential LIVs that are not predicted by AST or SR include tests on whether photons decay or split at superluminal speeds \cite{59}, and the SOL varying based on energy levels \cite{58}.  AST predicts that light always travels at the isotropic velocity $c$ within its PRF, and thus does not predict superluminal speeds for photons \cite{7, 22}.  Similarly, neither AST nor SR predicts that light speed is altered based on photon energy levels.  

\subsection{Category 2: Experiments that do not alter motion for AST} \label{section:4.2}
Experiments in category 2 test for PRFs that are external to the Earth.  The strategy used in these experiments is to maintain an experimental apparatus immobile on the Earth's surface while the Earth's rotation and orbital motion alters the velocity of the apparatus relative to a potential external PRF (Table 2).  These experiments measure for changes in experimental outcomes over time as the velocity relative to a potential external PRF changes (Fig. 3a).  However, this strategy does not test AST because the local PRF for AST is the ECI, which is an internal, not external, PRF for experiments on the Earth's surface.  An apparatus on the Earth's surface experiences constant rotational motion relative to the ECI (Fig. 3b).  Thus, AST predicts no changes in velocity over time for these types of experiments, which is consistent with the results.

\begin{table}[ht]
\footnotesize
\begin{flushleft}
\textbf{Table 2. Category 2: No changes in motion for AST over time.}
\end{flushleft}
\label{Table2}
\rowcolors{1}{}{lightgray}
\begin{tabular}{lll}%%%The number of columns has to be defined here
\hline 
Experiment type &Rotation &AST compatibility\\
\hline
\makecell[tl]{Pendulum motion of torsion spring\\with polarized electrons \cite{69}} {\hspace{12.17mm}} & Earth rotation & \makecell[tl]{Motion relative to ECI\\does not change over\\time (No AST motion) \cite{11}} \\

One-way maser frequency \cite{70} & Earth rotation & No AST motion \cite{11} \\

\makecell[tl]{Compton scattering of photons\\by high-energy electrons \cite{71}} {\hspace{17.3 mm}} & Earth rotation & No AST motion \cite{11} \\

\makecell[tl]{Splitting and recombining\\of electron wave packets \cite{30}} & Earth rotation & No AST motion \cite{11}
\vspace*{2pt}
\vspace*{3pt}
\\\hline
\end{tabular}
\end{table}%%%End of the table

	The experiments in category 2 include the Compton scattering of laser photons on high-energy electrons \cite{71}, the splitting and recombining of an electron wave packet \cite{30}, and one-way light paths to measure changes in frequency \cite{70}.  Since the experiments do not test AST, they do not distinguish between the simultaneity frameworks of AST and SR.

\subsection{Category 3: Stationary-frame relativistic effects}\label{section:4.3}
The experiments in category 3 exclusively analyze relativistic effects from the ``stationary'' perspective (Table 3).  From the ``stationary'' perspective, ALT and the LT share the same kinematics and make the same predictions (see Subsec. \ref{section:2.1}).  Thus, these experiments do not distinguish between their different simultaneity frameworks.  

\begin{table}[ht]
\footnotesize
\begin{flushleft}
\textbf{Table 3. Category 3: Stationary-frame relativistic effects that are shared by AST and SR.}
\end{flushleft}
\label{Table3}
\rowcolors{1}{}{lightgray}
\begin{tabular}{lll}%%%The number of columns has to be defined here
\hline
Experiment type &Rotation &AST compatibility\\
\hline
\makecell[tl]{Ives--Stilwell TD of hydrogen ions\\measured with transverse Doppler effect \cite{72}} {\hspace{8.7 mm}} & Earth rotation & ALT predicts TD \cite{7} \\

Pion lifetime  \cite{73} & Earth rotation & ALT predicts TD \cite{7} \\

\makecell[tl]{Mass of high-velocity subatomic particles \cite{74}} {\hspace{7.3 mm}} & Earth rotation & \makecell[tl]{AST predicts relativistic\\mass/energy \cite{20}} \\

\makecell[tl]{Relativistic energy of subatomic particles \cite{75}} & Earth rotation & \makecell[tl]{AST predicts relativistic\\mass/energy \cite{20}} \\

\makecell[tl]{Relativistic Doppler shift in parallel-\\antiparallel atomic resonance \cite{76}} {\hspace{16 mm}} & Earth rotation & \makecell[tl]{ALT predicts relativistic\\Doppler effect \cite{25, 26}} \\

\makecell[tl]{One-way SOL between rocket and Earth \cite{35}} & \makecell[tl]{Rocket motion\\Earth rotation} & \makecell[tl]{ALT – SOL independent\\of motion of source \cite{22}} \\

\makecell[tl]{SOL from binary stars is independent\\of the motion of its source \cite{77}} {\hspace{14.6 mm}} & None & \makecell[tl]{ALT – SOL independent\\of motion of source \cite{22}} \\

Atmospheric muon lifetimes \cite{78} & None & ALT predicts TD \cite{7} \\

Muon lifetime in circular motion \cite{16} & Circular motion & ALT predicts TD \cite{7} \\
\vspace*{-10 pt}
\\\hline
\end{tabular}
\end{table}%%%End of the table

In these experiments, the ``stationary''-frame is either the ECI, which is the local PRF for AST, or the ``laboratory'' frame.  The resolution of many experiments is not sufficient to distinguish between the ECI and the laboratory frame.  For example, analysis of the lifetimes of atmospheric muons that are traveling at 0.995$c$ \cite{78} cannot distinguish whether the ``stationary'' frame is the ECI or the Earth's surface, which is rotating relative to the ECI at only $\sim$1.5 x 10\textsuperscript{-6}$c$.  However, experiments with sufficient resolution show that the ECI is the correct ``stationary'' frame, e.g. see Ref. 35.

The experiments in category 3 include measurements of TD for atomic and subatomic particles \cite{78, 16}, including the Ives-Stilwell experiment \cite{72}, as well as relativistic mass/energy for high-velocity subatomic particles \cite{74}.  Other experiments probe additional LT and ALT shared relativistic effects, including that the SOL is independent of the motion of its source \cite{77}, and the relativistic Doppler effect \cite{76}.  The results of these experiments are compatible with the identical ``stationary''-frame predictions of AST and SR.

\subsection{Category 4: Velocity-invariant null results }\label{section:4.4}
The experiments in category 4 give null results, i.e. results that are not altered by changes in velocity (Table 4).  The null results arise from kinematics that are shared by the LT and ALT (see Subsec. \ref{section:2.1}).  

\begin{table}[ht]
\footnotesize
\begin{flushleft}
\textbf{Table 4. Category 4: Velocity-invariant null results that are predicted by AST and SR.}
\end{flushleft}
\label{Table4}
\rowcolors{1}{}{lightgray}
\begin{tabular}{lll}%%%The number of columns has to be defined here
\hline
Experiment type &Rotation &AST compatibility\\
\hline
Constancy of two-way SOL \cite{79} & Earth rotation & Shared kinematics \cite{22} \\

\makecell[tl]{Frequency of two-photon absorption\\with a fast atomic beam \cite{80}} & Earth rotation & Shared kinematics \cite{81, 82} \\

\makecell[tl]{Michelson--Morley interferometry two-way\\light paths of equal length \cite{83}} {\hspace{1.77 mm}} & \makecell[tl]{Rotation apparatus\\Earth rotation} & Shared kinematics \cite{27} \\

\makecell[tl]{Kennedy--Thorndike interferometry\\two-way light paths of unequal length \cite{84}} & \makecell[tl]{Rotation apparatus\\Earth rotation} & Shared kinematics \cite{27} \\

\makecell[tl]{Optical resonator maintains standing\\optical wave during rotation \cite{85}} {\hspace{8.15 mm}} & \makecell[tl]{Rotation apparatus\\Earth rotation} & Shared kinematics \cite{11} \\

\makecell[tl]{M\"ossbauer effect absorption of Doppler-\\shifted gamma rays during rotation \cite{86}} & \makecell[tl]{Rotation apparatus\\Earth rotation} & Shared kinematics \cite{87, 88} \\

\makecell[tl]{No change in frequency for one-way\\light path during rotation \cite{89}} {\hspace{10 mm}} & \makecell[tl]{Rotation apparatus\\Earth rotation} & Shared kinematics \cite{11} \\

\makecell[tl]{No change interferometry for light in\\vacuum and glass during rotation \cite{90}} & \makecell[tl]{Rotation apparatus\\Earth rotation} & Shared kinematics \cite{91} \\

\makecell[tl]{No change interferometry for light in air\\and water during rotation \cite{92}} {\hspace{4 mm}} & \makecell[tl]{Rotation apparatus\\Earth rotation} & Shared kinematics \cite{93} \\

\makecell[tl]{No frequency change two masers arrayed\\in opposite directions during rotation \cite{94}} & \makecell[tl]{Rotation apparatus\\Earth rotation} & Shared kinematics \cite{94}
\vspace*{4pt}
\\\hline
\end{tabular}
\end{table}%%%End of the table

To understand how null results are generated from the shared kinematics, consider the Michelson--Morley experiment with one of the light paths parallel to the direction of motion \cite{83}.  From the ``stationary'' perspective, light travels a longer distance to reach the mirror in the forward direction (because the mirror is moving away from the oncoming light), and a shorter distance to reach the mirror in the backward direction (because the mirror is moving toward the oncoming light).  As the velocity increases, the light travel distance becomes longer in the forward direction, and shorter in the backwards direction (due to the increased velocity of the mirrors away from and toward the light in transit, respectively).  However, length contraction shortens both the forward and backward light paths, and this keeps the overall two-way light path the same length regardless of the velocity \cite{95}.  For the light path perpendicular to the direction of motion, the relativistic stellar aberration angle is required to properly angle the mirrors so that the light continues to be reflected between the mirrors despite the motion of the apparatus while the light is in transit between the mirrors \cite{11}.  These two relativistic effects are sufficient to generate a null result \cite{11}.  

Additional shared relativistic effects are required to produce null results in other experiments.  The constancy of the two-way SOL \cite{79} arises from length contraction and TD \cite{96}.  The null result in a Kennedy--Thorndike experiment \cite{84} arises from length contraction, TD, and the relativistic stellar aberration angle \cite{95, 11}.  The null result in an optical resonator experiment \cite{85} arises from length contraction, TD, the relativistic stellar aberration angle, and the relativistic Doppler effect \cite{11}. 

	Experiments that measure wavelength or frequency in one-way light paths also generate null results due to shared kinematics \cite{11}.  These include M\"ossbauer experiments \cite{86} and the analysis of one-way maser frequencies \cite{94}.  Since ALT and the LT share the kinematics that generate these null results, these experiments do not distinguish between their simultaneity frameworks.  

\subsection{Category 5: Alternate explanations}\label{section:4.5}
The experiments in category 5 focus on the spin-orbit interaction of elementary charged particles \cite{97, 98}, for which there are alternate theoretical explanations (Table 5).  In 1926, it was recognized that the predicted spin-interaction energy was too large by a factor of two to generate the observed anomalous Zeeman effect \cite{99}.  Thomas proposed that an electron in curvilinear motion undergoes rotation due to changes in velocity, which generates Thomas precession \cite{100}.  Thomas precession introduces a factor of 1/2 to the spin-interaction energy, which brings the predicted value in line with observations \cite{100}.  Thomas precession requires differential simultaneity \cite{27}. 

\begin{table}[ht]
\footnotesize
\begin{flushleft}
\textbf{Table 5. Category 5: Alternate explanations for spin-orbit interaction.}
\end{flushleft}
\label{Table5}
\rowcolors{1}{}{lightgray}
\begin{tabular}{lll}%%%The number of columns has to be defined here
\hline
Experiment type &Rotation &AST compatibility\\
\hline
\makecell[tl]{Anomalous Zeeman effect of magnetic\\field on spectral lines of electron \cite{101}} {\hspace{8.1 mm}} &\makecell[tl]{Curvilinear motion\\Earth rotation}&\makecell[tl]{Hidden momentum \cite{102, 103, 104}}\\

\makecell[tl]{Electron g-factor (dimensionless\\magnetic moment) \cite{98}} & \makecell[tl]{Curvilinear motion\\Earth rotation} & Hidden momentum \cite{102, 103, 104}\\

g-factor for muon in circular motion \cite{97} & Circular motion & Hidden momentum \cite{102, 103, 104} \\
\vspace*{-10pt}
\\\hline
\end{tabular}
\end{table}%%%End of the table

An alternate mechanism to obtain the 1/2 factor for the spin-interaction energy comes from the incorporation of ``hidden momentum'' \cite{102, 103, 104}.  Hidden momentum is the small amount of net mechanical momentum that is associated with the constituent parts of a system at rest \cite{105}.  Thomas precession is the historical and widely accepted explanation.  However, hidden momentum has the distinction of providing correct solutions in other contexts, including the force on the magnetic dipole \cite{106}, the Shockley--James paradox \cite{107}, Mansuripur's paradox \cite{108}, and the nonlocality of the Aharonov--Casher quantum effect \cite{106}.  Hidden momentum and Thomas precession cannot both be correct, as that would generate an erroneous value for the spin-orbit interaction \cite{109}.  Thus, either Thomas precession arises in the context of differential simultaneity, or hidden momentum functions in the absence of differential simultaneity.

\subsection{Category 6: Velocity-dependent rotational effects}\label{section:4.6}
The experiments in category 6 analyze velocity-dependent relativistic effects in rotating frames (Table 6).  Unlike the experiments in categories 1--5, these experiments can distinguish the simultaneity frameworks of SR and AST.  That is because SR and AST give different predictions for velocity-dependent relativistic effects when observations are made from within moving or rotating frames (see Sec. \ref{section:2}).  There are three subcategories of experiments.  

\begin{table}[!ht]
\footnotesize
\begin{flushleft}
\textbf{Table 6. Category 6: Directional relativistic effects predicted by AST but not SR.}
\end{flushleft}
\label{Table6}
\rowcolors{1}{}{lightgray}
\begin{tabular}{lll}%%%The number of columns has to be defined here
\hline
Experiment type &Rotation &AST compatibility\\
\hline
\makecell[tl]{Around-the-world flights\\demonstrating directional TD \cite{14, 15}} {\hspace{10.9 mm}} & \makecell[tl]{flights around Earth\\Earth rotation} & \makecell[tl]{ALT predicts \\ directional TD \cite{7}} {\hspace{4 mm}} \\

\makecell[tl]{Interferometry shows one-way\\SOL anisotropy on Earth surface \cite{33}} & Earth rotation & \makecell[tl]{ALT predicts one-way\\SOL anisotropy \cite{7}} \\

\makecell[tl]{Around-the-world Sagnac effect\\shows one-way SOL anisotropy \cite{34}} {\hspace{12.9 mm}} & Earth rotation & \makecell[tl]{ALT predicts\\Sagnac effect \cite{110}} {\hspace{4 mm}} \\

\makecell[tl]{Ring laser gyroscopes measure\\Earth rotation using Sagnac effect \cite{111}} & Earth rotation & \makecell[tl]{ALT predicts\\Sagnac effect \cite{110}} \\

\makecell[tl]{Phase shift neutron wave function\\detects Earth rotation \cite{112}} {\hspace{11.7 mm}} & Earth rotation & \makecell[tl]{ALT predicts phase\\shift Sagnac effect \cite{44}} \\

\makecell[tl]{Phase shift atomic wave function\\detects Earth rotation \cite{113}} & Earth rotation & \makecell[tl]{ALT predicts phase\\shift Sagnac effect \cite{44}}\vspace*{2pt}
\\\hline
\end{tabular}
\vspace*{-4pt}
\end{table}%%%End of the table

Experiments in the first subcategory demonstrate that one-way light speeds are anisotropic in rotating frames.  The one-way SOL was first demonstrated to be anisotropic on the Earth's surface in 1925 by Michelson and Gale \cite{33}.  A later GPS experiment demonstrated that light sent around the world in the direction of rotation (eastward) takes longer to reach a sending/receiving point than light sent in the counter-rotating direction (westward) \cite{34}.  The one-way SOL in the rotating frame can be calculated by dividing the distance in the rotating frame (i.e. the circumference of the Earth at a given latitude) by the times for light to traverse that distance.  The experimental results indicate that the one-way light speed in the rotating frame is slower than $c$ eastward and faster than $c$ westward. 

AST directly predicts the one-way and two-way light speeds that are observed on the Earth's surface \cite{32, 22, 44}.  The rotational form of ALT predicts the observed isotropic two-way SOL (\ref{eq:45}) as well as the anisotropic one-way SOL (\ref{eq:47}) that generates the observed conventional Sagnac effect (\ref{eq:47}).  In contrast, the predictions of the other three major rotational transformations do not match experimental observations.  The  Langevin metric \cite{114, 115} and Post transformation \cite{116} predict anisotropic two-way speeds of light and variant Sagnac effects that are not observed \cite{32}.  The LT/Franklin transformation predicts the isotropic two-way SOL (\ref{eq:44}) but gives a null Sagnac effect (\ref{eq:49}) that is not observed \cite{32}.  Thus, the rotational form of ALT is unique in accurately describing the observed isotropic two-way SOL, the anisotropic one-way SOL, and the conventional Sagnac effect.

Experiments in the second subcategory demonstrate that TD is directional.  In the Hafele and Keating experiment, airplanes carrying synchronized clocks were flown around the Earth in eastward and westward directions \cite{14, 15}.  The clocks were directly compared before and after the flights.  Despite the flight times around the world being roughly equivalent in the two directions \cite{117}, the clocks traveling eastward (which have a higher velocity relative to the ECI) experienced more TD than the clocks traveling westward \cite{14, 15}.

AST directly predicts the time contraction experienced by the westward-flying clocks in the Hafele and Keating experiment through the converse ALT relation (\ref{eq:17}).  In contrast, the converse LT relation (\ref{eq:30}) or the converse relation for the rotational form of the LT \cite{45} predict TD rather than time contraction.  Thus, SR does not have the mathematical framework to describe the observations from the rotating-frame perspective, while AST does.  A theoretical explanation to explain the observed directional TD based on the time gap hypothesis \cite{118} has been invalidated by experimental data \cite{32}.  Therefore, the observation of directional TD is compatible with AST but not SR, and this implies the absence of differential simultaneity.

Experiments in the third subcategory demonstrate that there are velocity-dependent changes in the phase shift of neutron and atomic wave functions based on the Earth's rotation \cite{112, 113}.  ALT predicts these Sagnac effects \cite{44}, while the Sagnac effect is not predicted by the LT (\ref{eq:49}) \cite{32}. 

Authors for two of the category 6 experiments proposed how SR could be compatible with the experimental results despite describing light speed anisotropy and directional relativistic effects.  In the around-the-world GPS experiment, the authors suggest that they have obtained direct evidence for differential simultaneity because their observed anisotropic time delay has the same magnitude as the RTO, which describes the offsetting of time with distance \cite{34}.  However, the authors apparently failed to appreciate that the RTO and the observed anisotropic timing have opposite signs, and therefore are not equivalent.  In the co-rotating direction, the RTO is $-2 \pi \omega r\textsuperscript{2}/c\textsuperscript{2}$ (see Fig. 2a), while the one-way SOL anisotropy is longer than isotropic timing by $+2 \pi \omega r\textsuperscript{2}/c\textsuperscript{2}$ (because the sender/receiver is rotating away from the oncoming light from the ``stationary'' perspective) \cite{1, 12}.  In fact, the absence of the RTO is what allows the one-way SOL to be anisotropic.  If the RTO were present, then its alteration of clock timing would exactly negate the anisotropic timing to make the one-way SOL isotropic.  Therefore, the authors' observation of one-way light speed anisotropy that is the magnitude of the RTO but of opposite sign implies the absence of the RTO and differential simultaneity.

The authors of the around-the-world GPS experiment proposed a clock synchronization scheme that makes clock times equivalent around the globe for a ``stationary'' observer, which is what is expected in the AST framework.  However, a scheme to synchronize spatially separated clocks relative to each other does not affect the time keeping of a single clock.  In their experiment, only a single clock at the emission/reception site was necessary to determine the timing for light propagation.  Thus, the clock synchronization scheme would not alter their observation of one-way light speed anisotropy in the rotating frame, i.e. that a single clock records a longer time for light signals to traverse the globe eastward and a shorter time westward.  

In another category 6 experimental paper, Hafele and Keating proposed that SR is not applicable to rotating frames, and therefore should only be applied from the ``stationary'' perspective \cite{14, 15}.  Their proposal bypasses the issue that the predictions of the LT/Franklin transformation from the rotating perspective do not match the observed directional relativistic effects.  However, it does not distinguish between AST and SR because they have equivalent predictions from the ``stationary'' perspective.  Therefore, these proposals for compatibility of SR with the category 6 experiments do not alter the fact that AST is fully compatible with those experiments from both the rotating and stationary perspectives.

\section{Discussion}\label{section:6}
This study demonstrates that the broad range of current relativistic experiments are compatible with AST.  Since AST does not encompass differential simultaneity, this implies that current relativistic experiments do not provide definitive evidence for differential simultaneity.  This conclusion holds regardless of one’s opinion on whether AST will ultimately be found to be valid or any perceived gaps in the theoretical underpinnings of AST.  Until there is experimental evidence that invalidates AST (and, as described here, AST is consistent with current experiments), then AST can be used as a contrasting transformation to show that differential simultaneity has not yet been definitively demonstrated.  

The comparison of ALT and the LT identifies the types of experiments that can distinguish the simultaneity framework.  Since ALT and the LT give identical predictions from the ``stationary'' perspective, it is necessary to obtain data from within ``moving'' or rotating frames to assess the simultaneity framework.  Within a ``moving'' or rotating frame, the simultaneity framework can be ascertained by the directionality of relativistic effects and the one-way SOL.  

As shown here, the majority of experimental types (categories 1--5) do not distinguish the simultaneity framework because they are compatible with both AST and SR.  In contrast, the category 6 experiments can distinguish the simultaneity framework because the observations are made within a rotating frame and assess the directionality of relativistic effects or the one-way SOL.  These experiments suggest the absence of differential simultaneity in rotating frames based on the observation of directional relativistic effects and anisotropic one-way light speeds, both of which are predicted by AST. 

The incompatibility of the predictions of SR with the anisotropic one-way light speed observed within rotating frames that gives rise to the Sagnac effect led Witte in 1914 to initially propose that SR is not applicable to rotational motion \cite{13}.  Multiple theoretical explanations for how SR could be compatible with the observed Sagnac effect (including the integration of co-moving linear IRFs, and the time gap hypothesis) have been invalidated by high-resolution optical data \cite{32}.  The incompatibility of SR with velocity-dependent rotational relativistic effects can be attributed to the absence of differential simultaneity.  Consistently, an analysis using high-resolution optical data has determined that there is zero offsetting of time with distance in rotating frames \cite{12}.  

In contrast to SR's apparent incompatibility with rotational motion, SR is considered to be applicable to linear, constant-velocity trajectories (linear IRFs) and to gravitational free fall \cite{1, 119}.  Thus, linear IRFs and gravitational free fall are the ideal settings to test for the presence or absence of differential simultaneity.  Such data has not yet been obtained.  Several issues have acted as impediments to the collection of this data.  One impediment is that until recently, the methodology for measuring the offsetting of time with distance had not been described.  An experimental framework that can accomplish this in the context of the Mansouri and Sexl test theory is now available \cite{11}.  There are impediments to the analysis of linear IRF-like and gravity free-fall contexts.  It is difficult to obtain a linear trajectory on the Earth's surface because the planet is rotating, and therefore all trajectories include rotational components.  A linear trajectory should be linear relative to the ECI and the HCI (the two relevant local reference frames).  This would presumably require a space-based experiment.  Impediments to testing gravitational free fall include the increased complexity of undertaking an experiment in the limited time and restricted space of a container in free fall.  However, these impediments can be overcome with sufficient resources.

There are three major experimental approaches that could provide definitive evidence for differential simultaneity.  The first approach is the analysis of the one-way SOL in any ``moving'' frame other than an AST-designated PRF.  Isotropic one-way light speed would provide definitive evidence for differential simultaneity, while anisotropic one-way light would suggest its absence.  Historically, there has been an impediment to this analysis.  Determining the linear one-way SOL usually involves timing with two clocks, one at each end of the light path.  It has been recognized that the method of clock synchronization can alter the one-way SOL that is measured \cite{7}.  However, recently, a strategy to synchronize clocks for a linear light path without affecting the outcome has been described \cite{11}.  This synchronization method has the limitation that it specifically tests the AST framework versus the SR framework.  However, because AST is the only experimentally viable form of ALT \cite{11, 32}, and ALT and the LT are the only potentially-viable Lorentz-like transformations \cite{27, 28}, such an experiment is worthwhile.

The second experimental approach that could provide definitive evidence for differential simultaneity would be the demonstration of reciprocal relativistic effects between IRFs.  While directional relativistic effects have been documented in rotating frames (see Subsec. \ref{section:4.6}), there is a complete absence of evidence for reciprocal velocity-dependent relativistic effects.  Such an experiment could be undertaken by a space-based observer (in linear motion) assessing relativistic effects in the ECI reference frame.  The observation of reciprocal relativistic effects in the ECI frame (e.g. TD and increases in relativistic mass/energy) would provide evidence for differential simultaneity.  Conversely, observing directional relativistic effects in the ECI frame (e.g. time contraction and decreased relativistic mass/energy) would imply the absence of differential simultaneity.

The third experimental approach would be to demonstrate the conventional Maxwell's equations.  Maxwell's equations implicitly incorporate SR and differential simultaneity \cite{120}.  Maxwell's equations are modified in the ALT framework \cite{22, 121, 122, 123, 124, 125, 126, 127, 128}.  However, the differences in the predications of the ALT-modified Maxwell's equations and the conventional equations are so small that they cannot be distinguished by current experiments \cite{121, 122, 128}.  Directed experiments to distinguish between these small differences have not been carried out.  However, one can envision a scenario in which future experiments assess Maxwell's equations in the context of rotational motion, and the experiments observe the directional relativistic effects predicted by the ALT-modified equations.  Such a result would be anticipated by the observation that AST has to date correctly predicted all velocity-dependent directional relativistic effects in rotating frames \cite{32}.  In contrast, there are no examples of velocity-dependent reciprocal relativistic effects in rotating frames as would be implied by the observation of the conventional Maxwell's equations.  Thus, Maxwell's equations would probably also need to be analyzed in linear IRF-like or gravitational free-fall contexts to provide evidence for differential simultaneity.

The simultaneity framework has been a neglected area of research.  It is hoped that this study provides a roadmap for future experiments that explore the simultaneity framework in different contexts.  Given that the simultaneity framework defines how time and space interact, the answers from such studies are of fundamental importance.

\end{document}